\documentstyle[preprint,12pt,aps,epsf]{revtex}
\newcommand{\be}{\begin{equation}}
\newcommand{\ee}{\end{equation}}
\begin{document}
\draft

\title{Excitons in C$_{60}$ studied by
       Temperature Dependent Optical Second-Harmonic Generation}
\author{A.-M. Janner, R. Eder, B. Koopmans, H. T. Jonkman and G. A. Sawatzky}
\address{
Laboratory of Applied and Solid State Physics, Materials Science Centre\\
University of Groningen, Nijenborgh 4, 9747 AG Groningen, The Netherlands}
\date{\today}
\maketitle
\begin{abstract}
The electric dipole forbidden
$^{1}$T$_{1g}$ excitonic state of solid
C$_{60}$
at $\hbar \omega$=1.81 eV can be probed with a Second-Harmonic Generation
(SHG) experiment \cite{koopmans-b}.
We show that the SHG line shape depends strongly on the degree of
rotational order. We observe a splitting into
two peaks below the rotational ordering phase transition temperature
of 260 K.
The origin of this splitting is discussed in terms of
a possible Jahn-Teller effect, a possible
Davydov splitting due to the four molecules per unit cell in
the low temperature phase, and a mixing of the nearly degenerate
$^1$T$_{1g}$ and $^1$G$_g$ free molecule states because of the
lower symmetry in the solid. The exciton band structure is
calculated with a charge transfer mediated propagation
mechanism as suggested by Lof {\em et al.} \cite{lof} and with
one-electron (-hole) transfer integrals determined from band
structure calculations. Comparison with our experimental SHG data leads
to a reasonable agreement and shows that a mixing of $^1$T$_{1g}$
and $^1$G$_g$ states may explain the splitting at low  temperature.
\end{abstract}

\section{Introduction}

In spite of the large amount of research done on C$_{60}$ and its
compounds since its discovery \cite{kroto}, the electronic structure and the
importance of solid state band structure effects remain controversial.
Solid C$_{60}$ seems to exhibit a dualistic behaviour. On one
hand it behaves like a molecular crystal in which the molecular
properties (like the vibrational states and electronic excitations) are only
weakly perturbed by the crystal symmetry, but on the other hand
it behaves like a semiconductor, with a moderate
(2.3 - 2.6 eV  \cite{lof,weaver})
band gap, which can be electron doped resulting in low energy impurity states
and band widths of about 0.6 eV \cite{saito,shirley}.
Also quite different from typical molecular crystals is that C$_{60}$ forms
ionic compounds, which in some cases exhibit metallic and even superconducting
\cite{hebert}
behaviour, clearly demonstrating the importance of one-electron band
formation. With regard to the properties of these compounds they are
reminiscent
of the charge transfer type of molecular solids like the much studied
TCNQ salts except that the C$_{60}$ compounds usually
show 3-dimensional behaviour rather than the 1- or 2-dimensional
behaviour exhibited by the charge transfer molecular solids. This, obviously,
is due to the spherical rather than linear or planar structure of
the molecule. In
this paper we present evidence that in pure C$_{60}$ also the excitons
exhibit this dualistic behaviour. The energies of the excitonic states are
close to those of the gas phase molecule emphasizing the molecular
characteristics, but the propagation of the exciton results in abnormally
large excitonic band widths and mixings of different multiplets
for a molecular solid. This can be explained
within a one- and two-particle band structure theory.

It is well established that the strong delocalization of the
p$_{\pi}$-electron network (as is also the case on a C$_{60}$ molecule)
can result in strong nonlinear
optical effects \cite{poly-nlo}. Koopmans
{\em et al.} \cite{koopmans-b,koopmans-res} have shown that the SH
signal is very strong due to a double resonance
if the primary energy is tuned to the $^1$T$_{1g}$
excitonic state at 1.81 eV. This provides a possibility to study the
excitons inside the electronic band gap, in particular the exciton
band width, the band splitting
due to crystallographic phase transitions, and the mixing of multiplets
due to the crystal symmetry.
To facilitate this study we have developed a theory for the exciton
splittings and dispersions based on the molecular multiplet splittings
and solid state effects arising from a charge transfer mechanism for the
exciton propagation \cite{lof,robert}. The one-electron (-hole)
hopping integrals required for this are obtained from a tight-binding
fit to the LDA band structure of C$_{60}$ as given by Satpathy
{\em et al.} \cite{satpathy}. We show that reasonable agreement can be
obtained with the experimental SH line shape with only one adjustable
parameter,
namely the $^1$T$_{1g}$ - $^1$G$_g$ molecular multiplet splitting.

\section{Experimental}

C$_{60}$ with a purity better than 99.99\% was evaporated from a
Knudsen cell onto
a substrate at UHV pressures below 4$\cdot$10$^{-9}$ mbar. As substrates we
used fused quartz or at
low temperature MgO, a good thermal conductor.

For the SH experiments a Nd:YAG laser
was used to pump a dye laser, producing 7ns pulses with an energy of
approximately 6 mJ/pulse and a repetition rate of 10 Hz. The fundamental
frequency
was scanned in the range $\hbar \omega$= 1.7...2.0 eV. The SH intensities were
calibrated by using a reference quartz crystal in a transmission
geometry, carefully tuned to a Maker fringe optimum by fine-tuning the
frequency, and corrected for changes of the coherence length
in the quartz crystal as function of the photon energy.
All SH experiments are performed at a fixed angle of incidence
(45$^{\circ}$ to the surface normal) and the specular reflected SH signal was
detected.

The
SHG of thin C$_{60}$ films exhibit complicated thickness- (and through the
dispersion also frequency-) dependent interference phenomena.
We showed before \cite{koopmans-b} that for a
mixed$_{\mbox{in}}-$p$_{\mbox{out}}$ polarization (m-p) combination (mixed
means 50\% p and 50\% s polarized light) the SH interference
pattern exhibits a broad minimum for C$_{60}$ film thicknesses of
around 250 nm. Therefore we chose a m-p polarization combination and a
thickness of 250 nm for the measurements presented in this paper, so
that dispersive interference effects can be neglected.

The temperature dependent SHG experiments were performed using a He-flow
cryostat (4 - 500 K). The temperature was measured with a thermocouple
glued to the substrate. Possible effects of heating during the laser pulse
were examined by varying the laser power. We found that below 100 K the
temperature during the laser pulse was about 20-30 K
higher than the one measured with the thermocouple. At higher temperatures,
in particular around the rotational ordering phase transition
temperature (260 K), no heating by the laser
pulse was detected.

\section{Experimental results}

In Fig.1 the SH intensity measured at various temperatures is shown
as a function of the fundamental photon frequency ($\hbar \omega$).
Around room temperature we observe the resonance at
about 1.8 eV already previously reported
\cite{koopmans-b,koopmans-res}.
Notice for decreasing temperature
the strong enhancement of the SH intensity, the overall blue
shift and the temperature dependence in the line shape of the resonance.
Fig.2a and 2b show the temperature dependence of the zeroth
and first
moment of the spectrum corresponding to the integrated intensity and to
the mean
frequency, respectively. In both cases we see a strong temperature dependence
at the phase transition temperature of 260 K.
It is also just below
this temperature that we observe a splitting of the resonance into two peaks
with an intensity ratio of about 3:1. The splitting is
about 40 meV. The total width of the signal at the base of the line
is approximately 100 meV, which is very large for an exciton
band width of a molecular crystal, as discussed below.

\section{Discussion}

Looking at the data in Fig.1 and 2, there are three main features to
be explained: $(i)$ the splitting of the signal
below the phase transition; $(ii)$ the line width larger than
expected for an electric dipole forbidden transition in a
molecular crystal, and $(iii)$ the overall strong temperature
dependence of the SH intensity and line shape.

We propose three possible mechanism for the splitting, namely
a Jahn-Teller effect, a Davydov splitting and a
mixing of the electronic molecular states.
A {\em Jahn-Teller splitting} can probably be discarded
since it is expected to be at most 12 meV for the
singlet states as determined by Wang {\em et al.} \cite{yu}.
Before discussing the other two possibilities we briefly review the
basic ideas involving the propagation of excitations and their observation
by SHG.

For a schematic picture of the double
resonant SHG process observed in this energy range we refer to Fig.1
of ref \cite{koopmans-b}. The three level diagram consists of
a magnetic dipole transition from the
molecular ground state to the $^1$T$_{1g}$ excited state (involving a h$_{u}
\rightarrow$ t$_{1u}$ single-electron transition), followed by an
electric dipole transition to a
$^1$T$_{1u}$ state at about 3.6 eV (h$_g \rightarrow$ h$_u$),
and finally an electric dipole transition back to the ground state
(t$_{1u} \rightarrow$ h$_g$).
Linear optical experiments exhibit a strong electric
dipole allowed transition at 3.56 eV with a half width at half maximum of
0.23 eV. Since this width is much larger than the one observed
in our experiment (0.06 eV) we concluded that the sharp features in
the SHG spectrum must be related to the
intermediate $^1$T$_{1g}$ exciton state \cite{k-thesis,k-saltlake}. This
difference
in width can easily be understood by comparing the intramolecular excitations,
where the electron and the hole are bound,
with the intermolecular electron-hole excitations. The latter determine the
conductivity gap involving dissociated electron-hole states. As measured by
photoconductivity \cite{photocond}, or by
combined photoelectron and inverse-photoelectron spectroscopy \cite{lof}
this gap is 2.3 eV.
In Fig.3 we show the energy level scheme of the intramolecular
excitonic excitations (on the left hand side) and the solid state
intermolecular band gap excitations (on the right hand side).
The molecular $^1$T$_{1u}$ state at 3.6 eV is well
inside the intermolecular electron-hole continuum and will decay into this
with a hopping integral comparable to the one-electron (-hole) band width.
The $^1$T$_{1g}$, however, is an electron-hole bound state, inside the
band gap,
and will therefore have a long lifetime. The extra energy required to
dissociate the electron-hole pair of the exciton (i.e. the exciton binding
energy) is directly related to the onsite Coulomb interaction measured to be
about 1.3 - 1.6 eV by Lof {\em et al.} \cite{lof}. Since the $^1$T$_{1g}$
exciton is bound and the transition to the ground state is electric dipole
forbidden, we expect it to be very long lived and we would
expect a very small
exciton dispersion by conventional optical dipole - optical dipole
intermolecular propagation. The still quite large total width of more than
100 meV is
therefore difficult to understand in the limit of a
molecular solid. It is, however, well known that band structure effects
in C$_{60}$ are not neglegible. Lof {\em et al.} \cite{lof} already
suggested a propagation mechanism which could lead to a substantial
exciton dispersion. A similar mechanism involving virtual charge transfer
states was previously suggested by Choi {\em et al.} \cite{silbey}
to explain the dispersional width of optically forbidden excitons in
molecular crystals.

This propagation mechanism is shown pictorially in Fig.4 which demonstrates
how an electron-hole pair on site $i$ can propagate
to a neighbouring site $j$ via a virtual excited intermediate state.
This intermediate nearest-neighbour (charge transfer) state, in which
the electron is on site $i$ and the hole on a nearest-neighbour
site $j$ (or vice versa), is
at an energy $U - V$ (the difference between the onsite Coulomb interaction
and the nearest-neighbour Coulomb repulsion
\cite{bluhwiler,comment-m}) higher than the exciton ground state energy.
The net effective
exciton hopping integral is given from perturbation theory by:
\be
T^{\mbox{exciton}} = \frac{2 t_e t_h}{U-V}
\ee
where $t_{e(h)}$ are the average single electron (hole) nearest-neighbour
hopping
integrals.
It should be noticed that the same electron and hole hopping integrals
are involved for the red shift of an exciton energy in the solid
relative to the corresponding
one in the gas phase.
The red shift is then
given in first perturbation theory by:
\be
\Delta E = K (\frac{t^2_e + t^2_h}{U-V})
\ee
where $K$ is a geometrical factor related to the symmetry of orbitals and
the nearest-neighbour coordination number. As an example, one has
$K$ = 12 \cite{auger} for a totally symmetrical one-electron
and one-hole orbital in a FCC lattice.
To estimate the red shift of the $^1$T$_{1g}$ state we look at the
isolated C$_{60}$ molecule where Gasyna {\em et al.} \cite{gasyna} found the
$^1$T$_{1g}$ at about 1.92 eV. The same value has been assigned to the
$^1$T$_{1g}$ state by Negri {\em et al.} \cite{negri} using the absorption
spectra of C$_{60}$ in n-hexane solution of Leach {\em et al.} \cite{leach}.
Koopmans
{\em et al.} found the $^1$T$_{1g}$ in solid C$_{60}$ resonant at 1.81 eV.
This means that in the solid state this exciton is red shifted by 110 meV,
which is in the same order as the exciton band width. This supports
the above mentioned exciton hopping mechanism. Detailed calculations
of the one-electron (-hole) hopping integrals (discussed further on)
show that only nearest-neighbour hopping have to be
considered \cite{robert}.
Therefore, this exciton hopping mechanism does not destroy the
molecular Frenkel character of the exciton.

The above described mechanism for the dispersion of a Frenkel exciton
is analogous
to the charge transfer mechanism proposed by Lof {\em et al.} \cite{lof}
and to the mechanism
originally used to describe excitons in molecular charge transfer salts
\cite{silbey}.
It is also very similar to the so called superexchange mechanism used to
describe Frenkel d-d excitons in 3d transition metal compounds
\cite{allen,sugano}.

Assuming that the exciton band width is primarily due to such a
dispersional width we look again at the temperature dependence of
the line shape.
First of all we might have expected to see only the zero quasi momentum
({\bf k} = 0) exciton because of the long optical wavelength.
At high temperatures, however, the molecules are rapidly rotating resulting
in dynamic orientational disorder which will cause a break down
of the translational lattice symmetry and of the
$\Delta${\bf k} = 0 selection rule. In the extreme case we would expect to
see just the total exciton density of states as we believe is indeed the
case at high temperatures.

Upon lowering the temperature below the phase transition at 260 K the
rotations are strongly reduced, leading to a decrease of dynamic disorder and
therefore to approach the $\Delta${\bf k} = 0 selection rule.
Since {\bf k} then becomes a good quantum number we will see
only the exciton states with {\bf k} vectors close to $\Gamma$, the center
of the Brillouin Zone.

In the
low temperature phase there are four molecules per unit cell so that the
exciton band at the $\Gamma$ point can split up into two or more bands.
This splitting, called the {\em Davydov splitting}, which represents a
first possible explanation, will be of the order of the
exciton band width and is prominently due to the dependence of the exciton
transfer integral on the relative orientation of neighbouring molecules.

Recent two-photon excitation of C$_{60}$ single crystal at 4 K by
Muccini {\em et al.}
\cite{muccini} shows a band at 1.846 eV which is assigned to the same lowest
forbidden Frenkel exciton of $^1$T$_{1g}$ symmetry as discussed
in this paper. They also find a second band at higher
energy (1.873 eV). They discuss this second band in terms of
a crystal field effect and as a possible Davydov splitting. They
give an alternative assignment of the second band as being due to a
second forbidden electronic state. Indeed semi-empirical quantum-chemical
calculations \cite{negri} have shown that there are several closely
spaced forbidden states which lie in a narrow energy range \cite{muccini}.
The two-photon spectrum of Muccini {\em et al.} strongly resembles the low
temperature SH resonance in Fig.1. However, their two-photon absorption,
being a third-order nonlinear optical experiment, involves
other selection rules than our SHG experiment.

In order to get a more detailed understanding of the results, we carried
out the full
exciton calculation starting from the basic ideas described by equation
(1) and (2). The details will be published elsewhere \cite{robert}. Here we
restrict ourselve to briefly describe the ingredients of the calculations
and the results. In the full calculation the orbital degeneracy of the
t$_{1u}$ (3 fold) and h$_u$ (5 fold) must be taken into account so that
there are several electron and hole hopping integrals depending on
the orbital quantum numbers. Satpathy {\em et al.} \cite{satpathy}
have described how those can be obtained from one-particle band
structure calculation using a tight-binding fit. The
electron and hole hopping integrals are a function of the relative
orientation of the buckyballs. These integrals are completely determined
from a single fit to the band structure for a particular given
structure. Also we must take
into account the multiplet structure of the molecular excitations
due to the intramolecular Coulomb interaction as
described by Negri {\em et al.}. These multiplet splittings are not
very well known but can be obtained from optical or electron
energy loss data of the gas phase or in solution.
The effective exciton
transfer integrals are then a sum of products of electron and hole
transfer integrals divided by $U - V$. The degree to which each of the
electron and hole hopping integrals contribute to the dispersion of a
particular exciton is determined by the weight of
the electron-hole product function in the
particular excitonic state under consideration.
In addition to the broadening of the molecular multiplets into bands,
there is also a mixing of the various molecular exciton states
because of the lowering of symmetry in the crystal.

The only remaining parameters are $U - V$ and the
molecular multiplet splitting.
Conserning $U - V$ a rough estimate can be taken from the Auger data
of Br\"{u}hwiler {\em et al.} \cite{bluhwiler}: U $\simeq$
1.1 eV (0.2 eV is substracted because of the higher exciton
energy of a singlet), and V $\simeq$ 0.7 eV.
This leaves us with $U - V \simeq 0.35$ $\pm$ 0.2 eV.
An independent estimate for $U - V$ can also be obtained from the experimental
red shift as given in eq.(2). Taking $U - V$ = 0.35 eV we get a calculated
red shift comparable to the experimental observed one.
Concerning the multiplet splitting we will see below that all we need
for the present propose is a small $^1$T$_{1g}$ - $^1$G$_g$
splitting.

These exciton dispersion calculations, show
that the $^1$T$_{1g}$ band at the $\Gamma$ point splits up in
three $^1$T$_g$ bands, one $^1$A$_g$ and one $^1$E$_g$ band
(Fig.5a) \cite{robert}.
This can be expected from group theoretical arguments because of
the transition from the space group
Fm$\bar{3}$m of the high temperature phase to the space group
Pa$\bar{3}$ of the low temperature phase
\cite{harris,dresselh}.
The Davydov splitting is found to be about 30 meV,
which is close to the experimental splitting (40 meV) of the two peaks. The
calculation, however, predicts that more than 90\%  of the weight would be
in the
lowest $^1$T$_{g}$ band. This is inconsistent with our data!
Another possible explanation appears when all molecular multiplet
states and their mixing is included.

As already mentioned, the quantum-chemical calculations of Negri
{\em et al.} \cite{negri} show that the $^1$T$_{2g}$ and $^1$G$_g$ states
(in terms of states of isolated C$_{60}$ molecules
with icosahedral symmetry) are nearly
degenerated with the $^1$T$_{1g}$.
In the crystal, however, the point group symmetry is lower. This gives rise
to a mixing of the icosahedral electronic eigenstates (compare with
Table VIII of \cite{dresselh}).
When this {\em mixing} of the $^1$T$_{1g}$
and $^1$G$_g$ Bloch states is
taken into account in the exciton dispersion
calculations \cite{robert}, a second somewhat smaller peak arises at higher
energy. This is another possible explanation for the splitting.
Accordingly, the main peak at 1.826 eV is (in terms of molecular states)
a mixed state of
$^1$T$_{1g}$ with some $^1$G$_g$ character, and the second peak at
1.866 eV is a $^1$G$_g$ state with some $^1$T$_{1g}$ character. Since
in our SHG experiment we probe the magnetic dipole allowed transitions
\cite{note}, only the $^1$T$_{1g}$ component is visible.
This would explain the difference in intensity.
Fig.5 shows
the calculated spectrum with and without mixing in of the $^1$G$_g$ state. The
"mixed" curve agrees well with the experimental data (at the lowest
temperature).

Although the $^1$T$_{2g}$ state is also very close to the $^1$T$_{1g}$,
the calculations show that these do not mix, because the neighbouring
molecules do not have the required orientation for allowing a mixing
of the corresponding electronic orbitals.
Notice that for the $^1$T$_{1g}$ and $^1$G$_g$ mixing the same exciton
transfer integrals are involved as in the case of a Davydov splitting.
Because of these exciton hopping integrals, described by eq.(1),
such a large Davydov splitting (compared to common molecular crystals
where an electric dipole forbidden transition is considered) and a
$^1$T$_{1g}$ and $^1$G$_g$ mixing are possible.

The blue shift of the first moment of the SHG spectrum
(Fig.2b) for decreasing temperature most probably
has its origin in the orientational ordering,
which takes place at the phase transition temperature (260 K).
In the low temperature phase (T $<$ 260 K) the C$_{60}$ molecules
can only jump between two equilibrium positions and
at T $<$ 100 K they are practically frozen in, whereas
in the high temperature phase (T $>$ 260 K) the buckyballs rotate freely
in all directions \cite{heiney,david}. Calculations of
the electron (hole) transfer integrals
for both phases show that the hopping integrals for the high temperature
phase are larger than those for the low temperature phase. This means
that in the low temperature phase, where a double carbon-carbon bond
faces a pentagon or hexagon, the exciton propagation
is less favourable, resulting in a narrowing of the band. Since we
are probing the $^1$T$_{g}$ state, which forms the bottom of the band, a
narrowing of the band gives rise to a blue shift of the $^1$T$_{g}$
state. The difference in magnitude of the low and high temperature
hopping integrals has its impact on still another process.
Eq.(2) gives the relation between the electron
(hole) hopping integrals and the red shift of a state in the solid compared
to the gas phase. Thus, we expect that at low temperature (where the
hopping integrals are smaller than those for the high temperature
phase), the $^1$T$_{1g}$ state is less red shifted than at high temperature.
This also resultes in a blue shift. Calculations, however, show that the
first process will be dominant in our observed blue shift.

What about the strong temperature dependence of the SH intensity?
Also this can be explained in terms of the dynamic rotational disorder.
SHG depends strongly on the retention of coherence in the $^1$T$_{1g}$
intermediate exciton state in a time scale determined by the excitation
transition matrix elements. The intensity of the SHG goes like
(N$_{\mbox{coh}})^2$ where N$_{\mbox{coh}}$ is the number of molecules
coherently (in phase) contributing to the signal. Rotational motion during
excitation results in dephasing of true oscillators so that
N$_{\mbox{coh}}$ is expected to decrease strongly with temperature.
This is a so called T$_2$ (i.e. dephasing time) like relaxation process.
One expects, therefore, that the rotation time of a buckyball is of
the same order as the time between the first and second transition.
We estimate the time of revolution for a freely rotating buckyball
at temperature T as:
$\tau_{rot} = 2 \pi \sqrt{I / 2k_BT}$,
where $I$=1$\cdot$10$^{-43}$ kgm$^2$ is the moment of inertia of
the buckyball.
For the time between the magnetic dipole transition and the
first electric dipole transition by
Fermi's Golden Rule gives
$\tau^{-1} \simeq r^2_{ball} \cdot G \cdot \rho_{final}$,
where $r^2_{ball}$ is the radius of the buckyball, $G$ is the
energy current of the laser pulse and $\rho_{final}$ is the
DOS of the final states (with one electron in the t$_{1u}$ (LUMO)
and one hole in the h$_g$ (HOMO $-$ 1), the second highest occupied molecular
orbital). We choose
$\rho_{final}$ = W$^{-1}$ where W$\sim$0.5 eV is the band
width of LUMO or HOMO $-$ 1. At T = 300 K we find $\tau_{rot}\simeq
\tau\simeq10^{-11}$ s. This means that, in the time spent between
the first and second transition a given buckyball can well performe
a full rotation. Since the rotation of the different molecules is
uncorrelated, this leads to strong (T$_2$) dephasing and decreasing of the
SH intensity.
Also a strong temperature dependence of the intensity can be found in
photoluminescence experiments. There, the increase in intensity
occurs at a somewhat lower temperature compared with the phase transition
and the interpretation is still controversial
\cite{pichler,graham,sauvajol,matus,schlaich}.

Liu {\em et al.} have also done temperature dependent SHG of C$_{60}$
films, and they also observe a jump in the SH intensity around the
orientational phase transition \cite{temp-china}. Their increase of
the SH intensity for decreasing temperature is much less than what we
observe. That is probably because they have done the SHG
experiment using a fixed frequency (1064 nm) which involves other
transitions well off the double resonance as found by Wilk {\em et al.}
\cite{dieter}.
Their SH intensity might also contain a change due to a shift of the
exciton states on going through the phase transition temperature of 260 K
as we have described above.

\section{Conclusions}

We have studied the dynamics of the $^1$T$_{1g}$ Frenkel exciton
at $\hbar\omega$ = 1.81 eV with temperature dependent SHG.
We find a very strong temperature dependence of this double
resonance. Its SH intensity increases strongly close to the phase
transition down to about 200 K. We explain this by correlating
the rotational disorder of the C$_{60}$ molecules to a T$_2$ dephasing
mechanism. Below the
rotational ordering phase transition the SH resonance
splits in two bands. Several ideas about what could be the cause of
this splitting are discussed.
Detailed exciton dispersion calculations taking into account
the full symmetry, multiplet structure and crystal structure, yield
large exciton dispersions with total band widths of about
100 meV and a Davydov splitting of the $^1$T$_{1g}$ state of
30 meV. The corresponding SH intensity is calculated to be
concentrated to more than 90\% in the lowest Davydov component
and this is not the observed behaviour. The
experimental data including the two component structure in the low
temperature phase is, however, very well described by the theory
if the full multiplet structure and mixing of the $^1$T$_{1g}$
and $^1$G$_g$ states due to the lowering of space group
symmetry is included. The experimental data together with the
theory support strongly the ideas that the excitons in solid
C$_{60}$ are Frenkel like but propagate via virtually
excited charge transfer states described by Lof {\em et al.}
\cite{lof}. This model is consistent with the total width, the
splitting below the phase transition, the red shift relative
to the gas phase and the blue shift with lowering temperature
of the $^1$T$_{1g}$ exciton state component measured in our SHG
experiment.

\section {Acknowledgements}

This investigation was supported by the Netherlands Foundation for
Fundamental Research on Matter (FOM) with financial support from the
Netherlands Organization for the Advancement of Pure Research (NWO).

\newpage

\noindent
{\bf FIGURE CAPTIONS}
\\

\noindent
1. The temperature dependent SHG of C$_{60}$ thin film using a m-p
polarization combination. The (low) temperatures are
not corrected for the heat induces by the laser.\\

\noindent
2. The temperature dependence of (a) the zeroth moment of the SHG spectrum
corresponding to the total intensity and of (b) the first moment
corresponding to the mean frequency ("averaged peak position").
Notice that the strongest
temperature dependence is at the phase transition temperature of 260 K
(T$_c$).\\

\noindent
3. Part of the singlet excitation spectrum of C$_{60}$ with on the
left side the intramolecular (on buckyball) excitations and on
the right side the solid state interband
excitations. The multiplet splitting of the optical forbidden
HOMO $\rightarrow$ LUMO excitations \cite{k-thesis,k-saltlake}
have an energy lower than the conductivity gap of 2.3 eV and are
therefore Frenkel
excitons. The (broad) arrows at the left show the three-level
diagram responsible for the 1.81 eV SHG resonance. The Frenkel excitons
can propagate via the nearest-neighbour charge transfer (C.T.) states,
which are $U - V$ higher in energy (see eq.(1)) \cite{bluhwiler,comment-m}.\\

\noindent
4. Schematic representation for the propagation of an exciton via
a charge transfer mediated mechanism; the electron hops with a
one-electron hopping intergral (t$_e$) to its nearest neighbour (n.n.).
This virtual charge transfer state is at an energy $U - V$
\cite{bluhwiler,comment-m} higher than the exciton state.
The hole then follows the electron
and the whole exciton has moved to the nearest-neighbour position. \\

\noindent
5. Comparison of the SHG data taken at the lowest temperature (27 K)
with the theoretical calculations for the low temperature phase
(Pa$\bar{3}$). The energy of the $^1$T$_{1g}$ state of the free
buckyball is taken to be 1.915 eV. The $^1$G$_g$ state is taken to be
at $+\infty$ (a) or degenerate with the $^1$T$_{1g}$ (b).
The values of $U-V$ are 0.3 eV (a) and 0.35 (b).
The bars represent the bands at the $\Gamma$ point;
their length correspondes to the degeneracy of the band,
3 fold ($^1$T$_g$), 2 fold ($^1$E$_g$) and non degenerate ($^1$A$_g$).


\begin{thebibliography}{99}
\bibitem{koopmans-b} B. Koopmans, A.-M. Janner, H.T. Jonkman, G.A. Sawatzky
and F. van der Woude, Phys. Rev. Lett. {\bf 71}, 3569 (1993).
\bibitem{lof} R.W. Lof, M.A. van Veenendaal, B. Koopmans, H.T. Jonkman
and G.A. Sawatzky, Phys. Rev. Lett. {\bf 68}, 3924 (1992).
\bibitem{kroto} H.W. Kroto, J.R. Heath, S.C. O'Brien, R.F. Curl and
R.E. Smalley, Nature {\bf 318}, 162 (1985).
\bibitem{weaver} P.J. Benning, J.L. Martins, J.H. Weaver, L.P.F. Chibante
and R.E. Smalley, Science {\bf 252}, 1417 (1991).
\bibitem{saito} S. Saito and A. Oshiyama, Phys. Rev. Lett. {\bf 66},
2637 (1991).
\bibitem{shirley} E.L. Shirley and S.G. Louie, Phys. Rev. Lett {\bf 71},
133 (1993).
\bibitem{hebert} A.F. Hebard, M.J. Rosseinsky, R.C. Haddon, D.W. Murphy,
S.H. Glarum, T.T.M. Palstra, A.P. Ramirez and A.R. Kortan, Nature {\bf 350},
600 (1991).
\bibitem{poly-nlo} J.L. Br\'{e}das and R. Silbey, {\em Conjugated Polymers. The
Novel Science and Technology of Highly Conducting and Nonlinear Optically
active Materials} (Kluwer, Dordrecht, 1991); A.J. Heeger, J. Orenstein and
D.R. Ulrich, {\em Nonlinear Optical Properties of Polymers},
(Materials Research
Society, Pittsburger, Pennsylvania, 1988); D.J. Williams, {\em Nonlinear
Optical
Properties of Organic and Polymeric Materials}, (American Chemical Society,
Washington D.C., 1983); J.L. Br\'{e}das and R.R. Chance, {\em Conjugated
Polymeric Materials: Opportiunities in Electronics, Optoelectronics and
Molecular Electroic}, (Kluwer, Dordrecht, 1990).
\bibitem{koopmans-res} B. Koopmans, A. Anema, H.T. Jonkman, G.A. Sawatzky
and F. van der Woude, Phys. Rev. B {\bf 48}, 2759 (1993).
\bibitem{robert} R. Eder {\em et al.} to be submitted.
\bibitem{satpathy} S. Satpathy, V.P. Antropov, O.K. Anderson, O. Jepsen,
O. Gunnarsson and A.I. Liechtenstein, Phys. Rev. B {\bf 46}, 1773 (1992).
\bibitem{yu} W.Z. Wang, C.L. Wang, A.R. Bishop, L. Yu and Z.B. Su,
Phys. Rev. B {\bf 51}, 10209 (1995).
\bibitem{k-thesis} B. Koopmans, {\em Interface and Bulk Contributions in
Optical Second-Harmonic Generation}, (Ph.D. Thesis, University of Groningen,
The Netherlands, 1993).
\bibitem{k-saltlake} B. Koopmans, A.-M. Janner, R. Guardini,
H.T. Jonkman, G.A. Sawatzky
and F. van der Woude, Mol. Cryst. Liq. Cryst. {\bf 256}, 299 (1994).
\bibitem{photocond} C.H. Lee, G. Yu, D. Moses, V.I. Srdanov, X. Wei and
Z.V. Vardeny, Phys. Rev. B {\bf 48}, 8506 (1993).
\bibitem{silbey} S.-I. Choi, J. Jortner, S.A. Rice and R. Silbey,
J. Chem. Phys. {\bf 41}, 3294 (1964).
\bibitem{bluhwiler} P.A. Br\"{u}hwiler, A.J. Maxwell, P. Rudolf, C.D. Gutleben,
B. W\"{a}stberg and N. Martensson, Phys. Rev. Lett. {\bf 71},
3721 (1993).
\bibitem{comment-m} M.B.J. Meinders, L.H. Tjeng and G.A. Sawatzky,
Phys. Rev. Lett. {\bf 73}, 2937 (1994); M. Meinders, J. van den Brink,
J. Lorenzana and G.A. Sawatzky, Phys. Rev. B, June 1995 (in press).
\bibitem{auger} G.A. Sawatzky and A. Lenselink, Phys. Rev. B {\bf 21},
1790 (1980).
\bibitem{gasyna} Z. Gasyna, P.N. Schatz, J.P. Hare, T.J. Dennis, H.W. Kroto,
R. Taylor and D.R.M. Walton, Chem. Phys. Lett. {\bf 183}, 283 (1991).
\bibitem{negri} F. Negri, G. Orlandi and F. Zerbetto, J. Chem. Phys.
{\bf 97}, 6496 (1992).
\bibitem{leach} S. Leach, M. Vervloet, A. Despr\`{e}s, E. Br\'{e}heret,
J.P. Hare, T.J. Dennis, H.W. Kroto, R. Taylor and D.R.M. Walton,
Chem. Phys. {\bf 160}, 451 (1992).
\bibitem{allen} J.W. Allen, R.M. Macfarlane and R.L. White, Phys. Rev.
{\bf 179}, 523 (1969).
\bibitem{sugano} S. Sugano, Y. Tanable and H. Kaminura, {\em Multiplets of
Transition Metal Ions in Crystals} (Academic, New York, 1970).
\bibitem{muccini} M. Muccini, R. Danieli, R. Zamboni, C. Taliani,
H. Mohn, W. M\"{u}ller and H.U. ter Meer, submitted to Chem. Phys. Lett.
(1995).
\bibitem{harris} T. Yildirim and A.B. Harris, Phys. Rev. B {\bf 46},
7878 (1992).
\bibitem{dresselh} G. Dresselhaus, M.S. Dresselhaus and P.C. Eklund,
Phys. Rev. B {\bf 45}, 6923 (1992).
\bibitem{note} In our SHG experiment we probe also electric quadrupole
allowed transitions, like the $^1$H$_g$ (see Fig.3). However, the
calculations of Negri {\em et al.} \cite{negri} and our calculations
\cite{robert} show that the $^1$H$_g$ state is about 300 meV above
the $^1$T$_{1g}$, far out of our detectable range.
\bibitem{heiney} P.A. Heiney, J.E. Fischer, A.R. McGhie, W.J. Romanow,
A.M. Denenstein, J.P. McCauley Jr., A.B. Smith III and D.E. Cox, Phys.
Rev. Lett. {\bf 66}, 2911 (1991).
\bibitem{david} W.I.F. David, R.M. Ibberson, T.J.S. Dennis, J.P.Hare
and K. Prassides, Europhys. Lett. {\bf 18}, 219 (1992).
\bibitem{pichler} K. Pichler, S. Graham, O.M. Gelsen, R.H. Friend, W.J.
Romanow, J.P. McCauley Jr., N. Coustel, J.E. Fisher and A.B. Smith III,
J. Phys.: Condens. Matter {\bf 3}, 9259 (1991).
\bibitem{graham} S. Graham, K. Pichler, R.H. Friend, W.J.
Romanow, J.P. McCauley jr., N. Coustel, J.E. Fisher and A.B. Smith III,
Synthetic Metals {\bf 49-50}, 531 (1992).
\bibitem{sauvajol} J.L. Sauvajol, Z. Hricha, N. Coustel, A. Zahab
and R. Arnar, J. Phys.: Condens. Matter {\bf 5}, 2045 (1993).
\bibitem{matus} M. Matus, H. Kuzmany and E. Sohmen, Phys. Rev. Lett.
{\bf 68}, 2822 (1992).
\bibitem{schlaich} H. Schlaich, M. Muccini, J. Feldmann, H. B\"{a}ssler,
E.O. G\"{o}bel, R. Zamboni, C. Taliani, J. Erxmeyer and A. Weidinger,
Chem. Phys. Lett. {\bf 236}, 135 (1995).
\bibitem{temp-china} Y. Liu, H. Jiang, W. Wang, Y. Li and J. Zheng,
Phys. Rev. B {\bf 50}, 4940 (1994).
\bibitem{dieter} D. Wilk, D. Johannsmann, C. Stanners and Y.R. Shen,
Phys. Rev. B {\bf 51}, 10057 (1995).

\end{thebibliography}
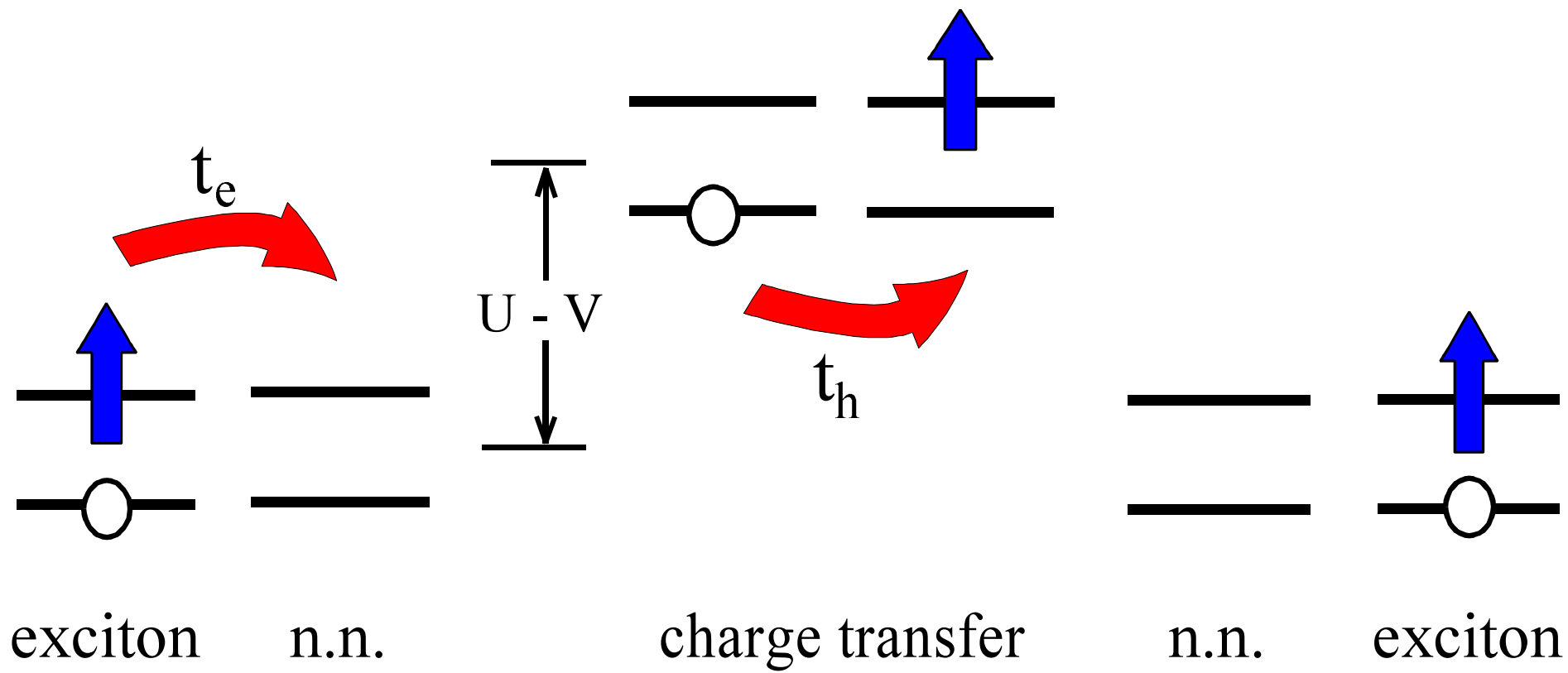
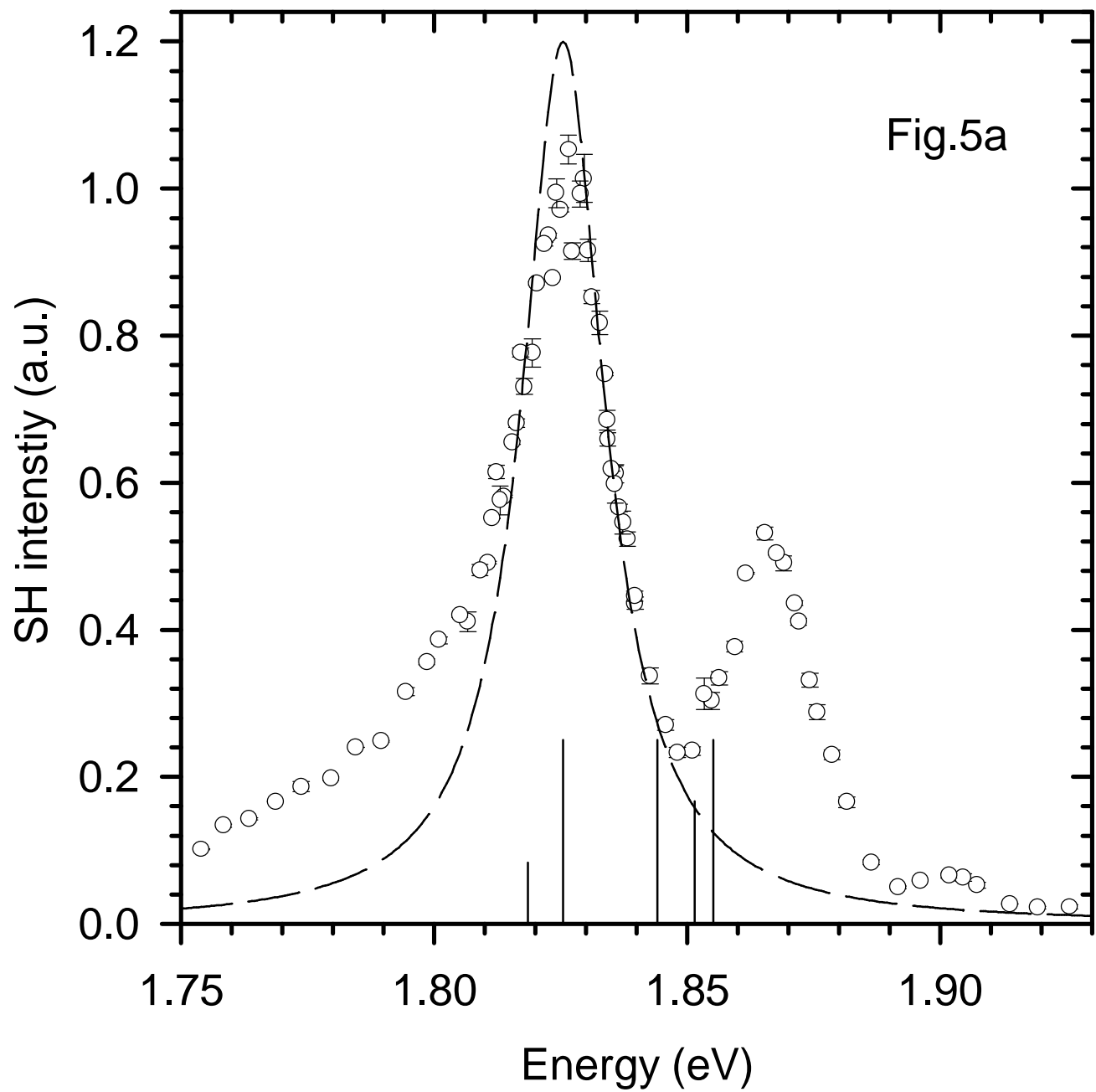
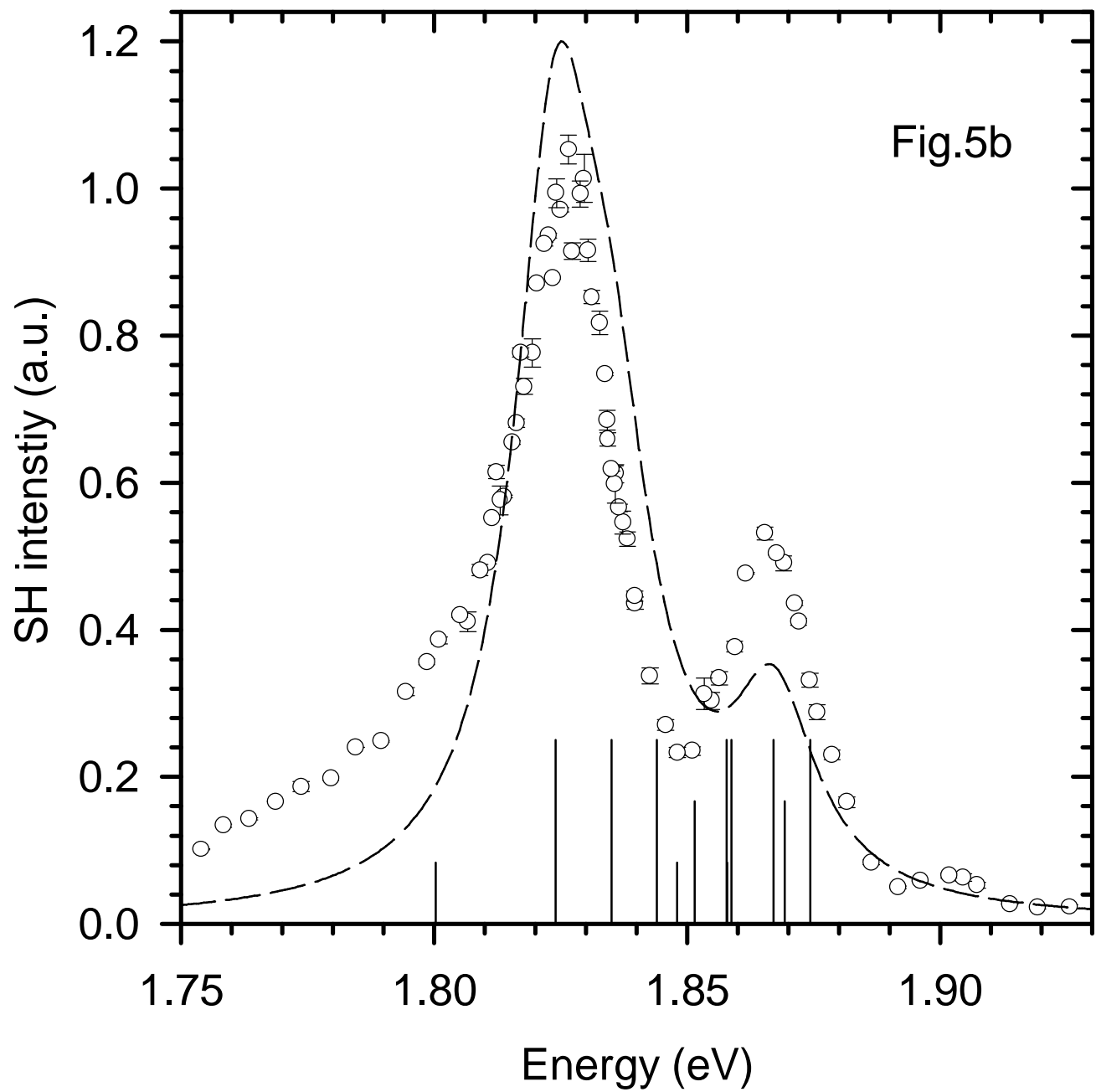
\end{document}